\documentclass[pra,showpacs,twocolumn]{revtex4}
\usepackage{graphicx}
\usepackage{amssymb}
\usepackage{rotating}

\newcommand{\dd}{\mbox{d}}

\newcommand{\nn}{\nonumber} 
\newcommand{\Eqref}[1]{Eq.~(\ref{#1})}
\newcommand{\ket}[1]{\left|{#1}\right\rangle}
\newcommand{\bra}[1]{\left\langle{#1}\right|}
\newcommand{\scalar}[2]{\left\langle\left.{#1}\right|{#2}\right\rangle}

\begin{document}

\title{Collisional decoherence of a tracer particle moving in one dimension}
\author{Ingo Kamleitner$^{1,2}$}
\affiliation{${}^1$Institut f\"ur Theory der Kondensierten Materie, Karlsruher Institut f\"ur Technologie, 76128 Karlsruhe, Germany}
\affiliation{${}^2$Centre for Quantum Computer Technology, Physics Department, Macquarie University, Sydney, NSW 2109, Australia}
\begin{abstract} 
	We study decoherence of the external degree of freedom of a tracer particle moving in a one dimensional dilute Boltzmann gas. We find that phase averaging is the dominant decoherence effect, rather than information exchange between tracer and gas particles. While a coherent superposition of two wave packets with different mean positions quickly turns into a mixed state, it is demonstrated that such superpositions of different momenta are robust to phase averaging, until the two wave packets acquire a different position due to the different velocity of each wave packet.
\end{abstract}
\maketitle

\section{Introduction\label{intro}}

The transition from quantum mechanics to classical physics is one of the most debated problems in the history of modern physics. In particular, the question arises why one can not observe macroscopic objects in a superposition of spatial distinct locations, despite the fact that all objects are made up of microscopic particles which indeed can be observed in such position superposition states. Several conceptually very different solution to this problem were proposed, as e.g.\ the theory of spontaneous localization~\cite{Ghirardi}, which modifies the Schr\"odinger equation by adding an incoherent part. A \emph{less drastic} approach  is the theory of environmentally induced decoherence~\cite{Zurek}. This assumes that the combined state of system and environment evolves according to Schr\"odinger's equation, but if only the system density operator is observed, it seems as if the coupling to the environment destroys the quantum feature that a system can be in a superposition of several distinct states, a process known as decoherence.

During the last two decades, there has been increasing interest in the engineering of large quantum systems, e.g.\ for quantum information processing. A major limitation to these efforts is posed by their fragileness to decoherence. Therefore, a detailed understanding of different decoherence processes is no more just an academic problem, but necessary for future quantum technologies.

A paradigm of environmentally induced decoherence is collisional decoherence, where the system of interest is a tracer particle, possibly macroscopic in size, which experiences random collisions with particles of a thermal reservoir. The colliding particles can be molecules, much as in Brownian motion, but one could also think of photons of the cosmic background radiation. Several authors put forward increasingly complicated master equations for a tracer particle in a thermal gas, first for an infinitely heavy tracer particle \cite{Joos,Gallis,Hornberger,Dodd}, and later for a tracer particle with finite mass \cite{Diosi,Vacchini,main,Hornberger2}. The latter were used to study collisional decoherence by applying quantum trajectory methods \cite{Breuer2,Hornberger4}. However, the validity of the single collision calculations used in the derivations of the respective master equations for tracer particle with finite mass was recently questioned \cite{Diosi2, Kamleitner}, and a consensus is still missing.

\begin{figure}[t]
	\includegraphics[width=\linewidth]{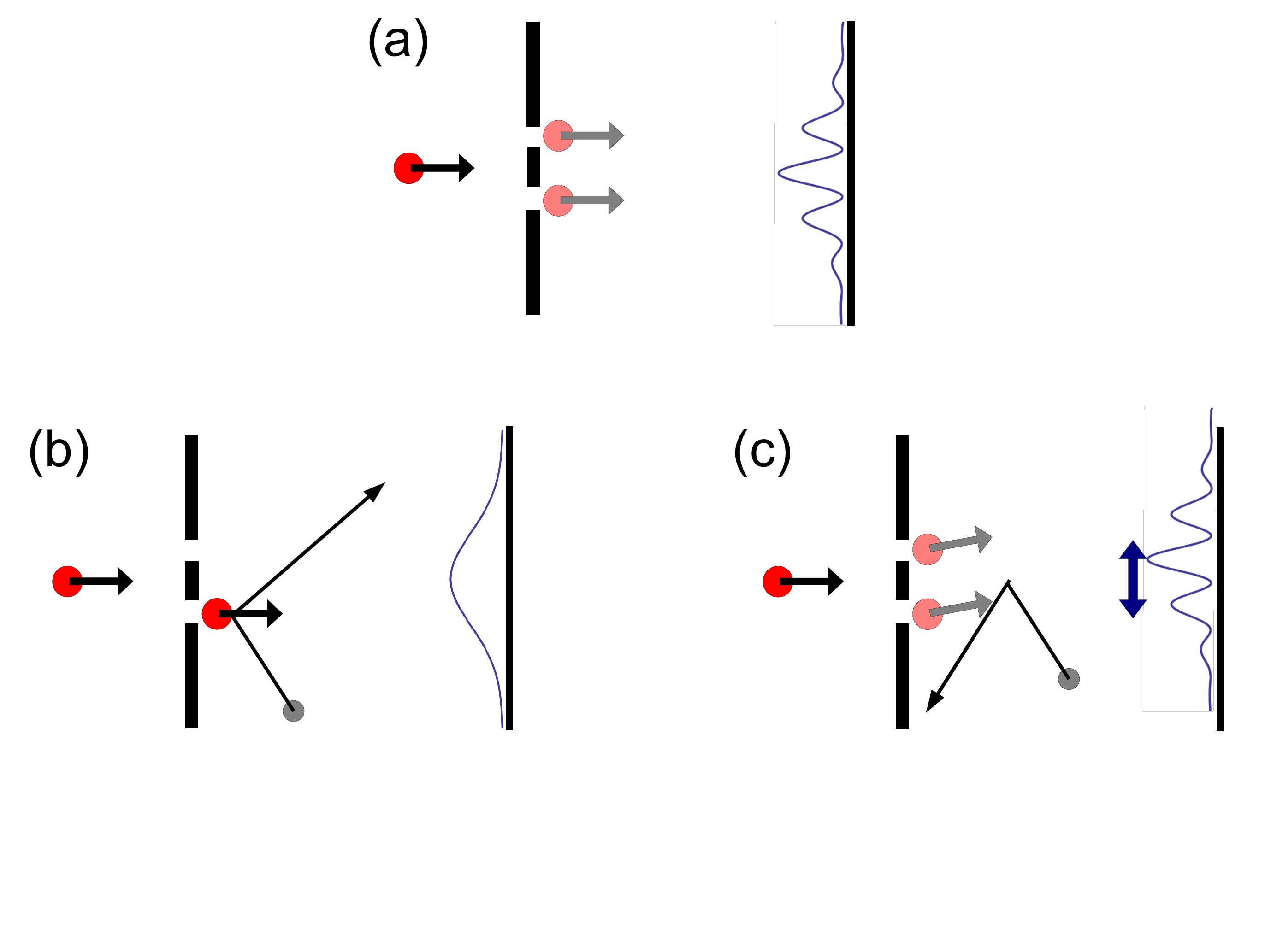}\vspace{-13mm}
\caption{\small (color online) (a): Undisturbed double slit experiment with screen in the far field (not to scale). (b): A colliding particle measures through which slit the tracer particle passes, removing the coherence pattern at the screen. (c): A colliding particle transfers momentum to the tracer particle, shifting the phase of the coherence pattern. If this shift is random, the coherences can not be observed anymore. One of the results of this paper is that process (c) is responsible for decoherence, rather than process (b).}
\label{fig}
\end{figure}

The mechanisms by which an environment can destroy measurable superpositions can roughly be divided into two categories. First, the environmental state can get entangled with the system state. This effectively delocalizes the relative phase of any superposition state of the system into the combined state of system and environment. After tracing out the environmental degrees of freedom, this leads to a reductions of the coherences of the system. We say, the environment \emph{measures} the system (see figure~\ref{fig}~(b)). In the second process, sometimes called phase averaging, the interaction with the environment changes the relative phase of the superposition state. If this phase change is random and different for each run of an experiment, then the coherence of the superposition state can no longer be observed (see figure~\ref{fig}~(c)). In some sense, phase averaging does not fundamentally destroy the coherences, but rather makes the relative phase unpredictable and therefore unmeasurable.

One should mention that neither mechanisms completely solves the problem of the quantum-classical transition because the \emph{measurement problem} remains~\cite{Zurek}. Nevertheless, both mechanisms can successfully describe the observed lack of coherences of macroscopic objects. However, it is the first decoherence mechanism which is mostly cited in connection with the quantum-classical transition, possibly because is appears to destroy coherences more fundamentally compared to phase averaging. This view is especially maintained within the topic of collisional decoherence \cite{Joos,Gallis,review}.

Although a colliding gas particle certainly carries away some information about the state of the tracer particle, we show in this article that the decoherence due to this information exchange is negligible compared to phase averaging. The latter arises because a collision adds a relative phase to a spatial superposition state, which depends on the momentum of the colliding gas particle and is therefore random. This article hence aims at a fundamental change of the understanding of the collisional decoherence process. 

Because we are interested in a general understanding of the collisional decoherence process, rather than in details depending on a particular interaction model, we use the simplest possible model. That is, we assume that the tracer particle as well as the gas particles only move in one dimension. Further, the gas is in a thermal state using Boltzmann statistics, and we apply the low density and high temperature limit. Within this limit, each collision is an independent event and we can neglect three particle collisions. The gas particles do not interact with each other (ideal gas), and their interaction potential with the tracer particle is of the hard core type, i.e.\ $V(\hat x-\hat x_g)= \lim_{a\to\infty}a\delta(\hat x-\hat x_g)$ where the index $g$ labels the gas particle.

Let us briefly review a single collision following reference~\cite{Kamleitner}. The effect of a collision on the tracer particle depends on the state of the colliding gas particle. Therefore, we start this discussion with a convenient convex decomposition of the thermal density operator of a gas particle. A particular useful convex decomposition was given by Hornberger and Sipe~\cite{Hornberger} in terms of Gaussian minimum uncertainty wave packets $\ket{x_g,p_g}_{\sigma_g}$ with
	\begin{eqnarray}
		\scalar{x_g'}{x_g,p_g}_{\sigma_g} &=& \frac{e^{-ix_gp_g/2\hbar}}{\sqrt{\sqrt\pi\sigma_g}} e^{ix'_gp_g/\hbar} e^{-(x_g-x'_g)^2/2\sigma_g^2} , \qquad 
	\end{eqnarray}
where $x_g$ and $p_g$ label the mean position and momentum of the wave packet, respectively, and $\sigma_g$ labels the position uncertainty. It was shown that the density operator can be written as
	\begin{eqnarray}
		\hat\rho_g &=& \frac{2\pi\hbar}{L}\mu_T(\hat p_g) \\
		&=& \int\frac{dx_g}L\int dp_g\, \mu_{T_{\sigma_g}}\!(p_g) \ket{x_g,p_g}_{\sigma_g}\!\!\bra{x_g,p_g}.\label{convex}
	\end{eqnarray}
Here, $\mu_T(p_g)=e^{-p_g^2/2m_gk_BT}/\sqrt{2\pi m_gk_BT}$ is the Maxwell-Boltzmann distribution, $L$ is a normalization length which is taken to infinity, and $T_{\sigma_g}=T-\frac{\hbar^2}{2m_gk_B\sigma_g^2}$. The reason for a lower temperature in the Maxwell-Boltzmann distribution in \Eqref{convex} is that part of the thermal energy of the gas particle has been transferred to being a contribution to the momentum uncertainty of the states $\ket{x_g,p_g}_{\sigma_g}$.

With \Eqref{convex} at hand, we can assume that every gas particle is in a minimum uncertainty state with position uncertainty $\sigma_g$, while the probability density for a particular combination of $x_g$ and $p_g$ is given by $n_g\mu_{T_{\sigma_g}}\!(p_g)$, where $n_g$ is the particle density of the gas.

For a complete collision of the gas particle wave packet with the tracer particle wave packet, it is required that the velocity uncertainty of the gas particle state $\ket{x_g,p_g}_{\sigma_g}$ is small compared to the relative velocity of the two colliding particles. It was shown in \cite{Kamleitner} that this is the case (at least for most gas particles) if $2m_gk_BT\sigma_g^2\gg\hbar^2$, and therefore we will choose $\sigma_g$ sufficiently large and approximate $\mu_{T_{\sigma_g}}\!(p_g)$ by $\mu_{T}(p_g)$ in the following. Furthermore, to avoid the discussion of three particle collisions, we require $\sigma_gn_g\ll1$. Therefore, the position uncertainty has to satisfy
	\begin{eqnarray}
		 \frac{\hbar}{\sqrt{m_gk_BT}} \,\;\ll &\sigma_g& \ll\;\, \frac1{n_g}, \label{limitation}
	\end{eqnarray}
which is generally possible in the high-temperature and low-density limit
	\begin{eqnarray}
		\frac{n_g\hbar}{\sqrt{m_gk_BT}}&\ll&1.
	\end{eqnarray}
Note that this limit must also be satisfied to use Boltzmann statistics to describe an ideal gas.

It was further shown in \cite{Kamleitner}, that, under the additional assumption of a slow (compared to the gas particles) tracer particle, the collision rate is
	\begin{eqnarray}
		R&=&n_g\sqrt{2k_BT}/\sqrt{\pi m_g}. \label{rate}
	\end{eqnarray}


It is well understood that if the tracer particle is initially (at time $t=0$) also in a minimum uncertainty state $\ket{x,p}_{\sigma}$, but with a position uncertainty $\sigma$ related to the gas particle's position uncertainty via their relative masses according to
	\begin{eqnarray}
		m\sigma^2 &=& m_g\sigma_g^2, \label{sigmarel}
	\end{eqnarray}
then a collision results in the remarkable simple product state~\cite{Schmuser,Kamleitner}
	\begin{eqnarray}
		U_g(t)\ket{\bar{x}_g,\bar{p}_g}_{\sigma_g}\otimes U_{}(t) \ket{\bar{x}_{},\bar{p}_{}}_\sigma. \label{gausstrafo}
	\end{eqnarray}
Here, $t$ is some time after the collision, and $U(t)$ is the free evolution operator of the single particle. The mean positions and momenta after the collision relate to the initial values according to
	\begin{eqnarray}
		\bar{x}_g = \frac{2 {x} -(1-\alpha) {x}_g}{1+\alpha}, && \bar{p}_g = \frac{2\alpha {p} -(1-\alpha) {p}_g}{1+\alpha}, \qquad \label{gasrelation}\\
		\bar{x}  = \frac{2\alpha {x}_g+(1-\alpha) {x} }{1+\alpha}, && \bar{p}  = \frac{2 {p}_g+(1-\alpha) {p} }{1+\alpha}, \label{tracerrelation}
	\end{eqnarray}
which are precisely the same relations as in a collision of classical particles. The relative mass is denoted by $\alpha=m_g/m$. The final state (\ref{gausstrafo}) was derived in ~\cite{Schmuser,Kamleitner} by two different methods, and can also easily be confirmed by using the scattering operator for a one dimensional hard core interaction, i.e.\ $\hat S \ket{p_g}\ket{p} = -\ket{\bar p_g}\ket{\bar p}$.

\begin{figure}[t]
	\includegraphics[width=\linewidth]{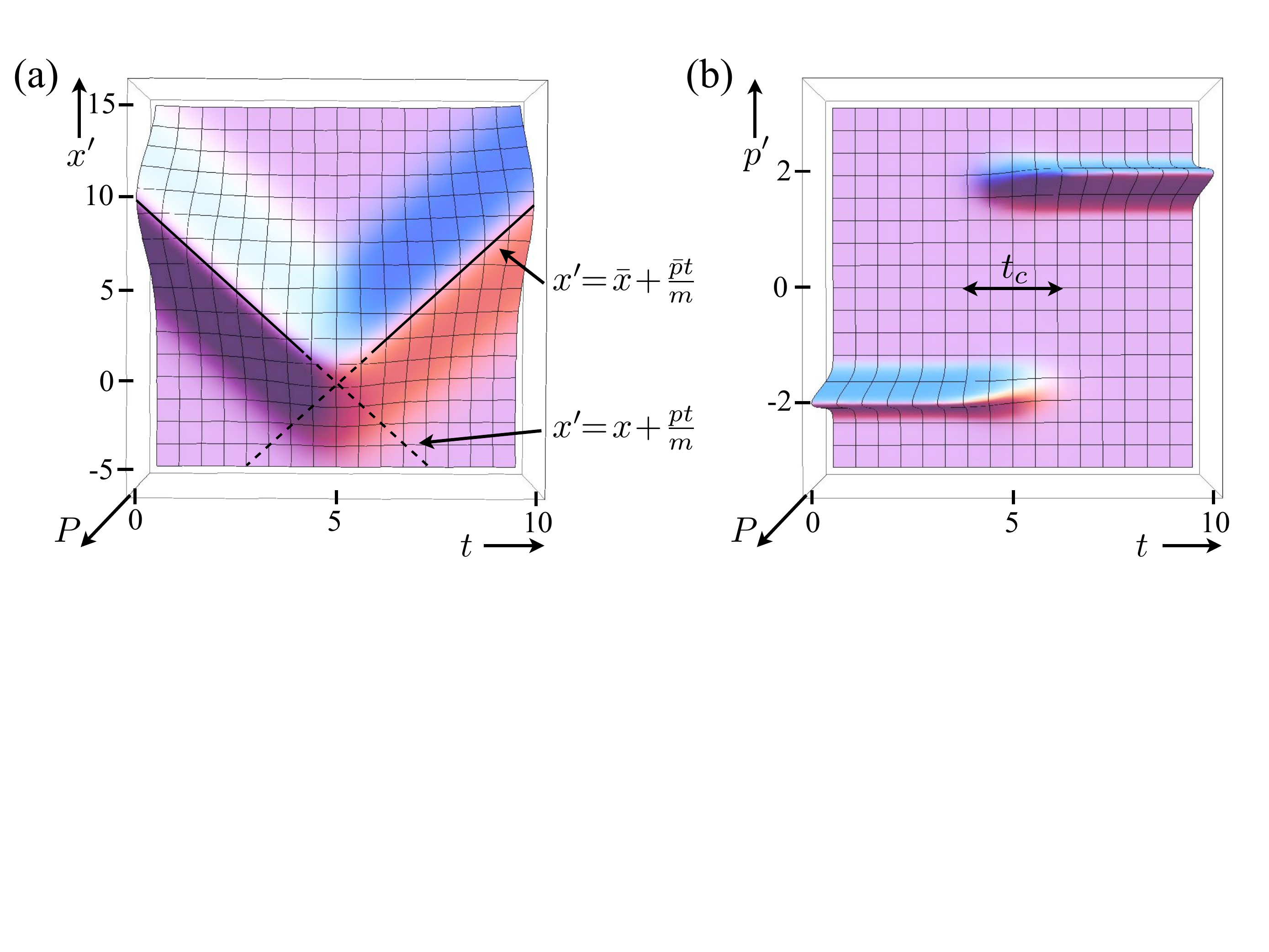}\vspace{-26mm}
\caption{\small (color online) Position (a) and momentum (b) probability distributions for the tracer particle during a collision in the reference frame of center of mass. While the position probability distribution ``flows", the momentum distribution ``jumps" from the initial value to the final one. The solid lines in (a) are the corresponding classical trajectories. The collision time $t_c$ is indicated in (b).}
\label{fig1}
\end{figure}

The reader should note that at no time the position probability distribution of either particle changes discontinuously, as might be suggested incorrectly by \Eqref{gasrelation} and \Eqref{tracerrelation}. The actual position probability distribution of the state~(\ref{gausstrafo}) also includes the free particle evolution operator, and in \cite{Kamleitner} it was shown that the evolution of the tracer particle in position representation is completely continuous, as can be seen in figure~\ref{fig1}~(a).

We should point out that in this work we neglect the possibility of incomplete collisions of the colliding wave packets. Therefore, our work is valid on a coarse grained time scale
	\begin{equation}
		t\gg t_c= \frac{2\sigma_g\sqrt{m_g}}{\sqrt{k_BT}}\gg \frac{\hbar}{k_BT}, \label{colltime}
	\end{equation}
where $t_c$ is the collision time defined in~\cite{Kamleitner} and shown in figure~\ref{fig1}~(b). For this very reason, we will use a finite time interval $(0,t)$ to study the decoherence process.

\section{Remarks about collisional decoherence\label{medi}}

Coherences of a quantum state are often described by the off-diagonals of the density matrix. This leads to the question: \emph{What basis should we use to examine decoherence?} The two bases which first come to mind are the position basis and the momentum basis, and indeed, these are the bases usually used in the literature~\cite{Joos,Gallis,Hornberger,review,Breuer2,Hornberger4}. Nevertheless, they are not without problems. 

If one uses e.g.\ the momentum basis, one would be interested in how long a superposition of the form $(\ket{p_a}+\ket{p_b})/\sqrt2$ survives, or, equivalently, how fast the coherences $\ket{p_a}\!\bra{p_b}$ decrease. There is a major problem in this sort of question: A momentum eigenstate (or any state which is very localized in momentum) is itself a coherent superposition of widely spread position eigenstates, and one should expect that these position coherences of each individual momentum eigenstate do vanish on the same time scale, or even faster, as the momentum coherences of interest! The same reasoning applies to using the position basis.

Our reservations are clearly related to the concept of a pointer basis~\cite{Zurek,lecture}. Pointer basis states should be fairly robust to decoherence. That is, if the system is prepared in one of these pointer states it will stay there for some time, whereas if the system is in a superposition of two pointer states, then the coherences typically decrease rapidly in time. These properties makes a pointer basis the basis of choice to study decoherence effects. Because a momentum eigenstate is highly unrobust to position decoherence, the momentum basis is not a good pointer basis. Similarly, a position eigenstate is unrobust to momentum decoherence, and therefore also not an appropriate pointer state.

To add some weight to our concern, we have a brief look at the decoherence rate found by~\cite{Breuer2} for an initial superposition $(\ket{p}+\ket{-p})/\sqrt2$ of momentum states, i.e. their equation~(67). Inserting the definitions~(27, 31, 32) of \cite{Breuer2} and assuming that the velocity of the tracer particle is small compared to an average gas particle, their equation~(67) leads to the decoherence rate $D_p= \frac{8\sqrt{2\pi}\sigma n_g}{3}\frac{p^2}{m^2}\sqrt{\frac{m_g}{k_BT}}$, where we changed the notations according to ours, and $\sigma$ is a constant scattering cross section. This formula strongly opposes physical intuition because it predicts a decrease of the decoherence rate upon increasing the temperature, despite the fact that an increase of the temperature leads to more powerful and more frequent collisions. 

Could we possibly single out a pointer basis from a measurement interpretation of a single collision? We showed in~\cite{Kamleitner} that a colliding gas particle $\ket{x_g,p_g}_{\sigma_g}$ performs a smeared out measurement on the tracer particle in the basis $\ket{x,p}_\sigma$, indicating that these states could be used as pointer basis. We know from ~\Eqref{convex}, that we have a choice in the width $\sigma_g$ of the gas particle states. Therefore, we can use minimum uncertainty states of any width $\sigma=\sqrt{m/m_g}\sigma_g$ as pointer basis, as long as $\sigma_g$ satisfies relation~(\ref{limitation}), which was required for the treatment of a single collision. 

As is stated in e.g.\ \cite{lecture}, there are still many open questions about the emergence of a pointer basis from the coupling to an environment. Nevertheless, because of the heuristic reasoning in the previous paragraph, and the lack of sensible alternatives, we will indeed discuss decoherence by using superpositions of Gaussian states $\ket{x_a,p_a}_\sigma+\ket{x_b,p_b}_\sigma$, commonly referred to as \emph{cat states} in the literature.  We should note, that~\cite{Hornberger4} superimposed several momentum eigenstates to produce a state similar to a cat state to study decoherence using quantum trajectory theory. This study was limited to the situation where the gas particle mass equals the tracer particle mass, and each collision leads to a complete loss of decoherence.

We will need the transformation of an initial (unnormalized) cat state $\ket{x_a,p_a}_\sigma+\ket{x_b,p_b}_\sigma$ of the tracer particle upon a collision with a gas particle in the state $\ket{x_g,p_g}_{\sigma_g}$. By using the linearity of quantum mechanics as well as Eqn.\ (\ref{gausstrafo})-(\ref{tracerrelation}), and after tracing out the gas particle, we find for the density matrix of the tracer particle after a collision
	\begin{eqnarray}
		\hat{\bar\rho} (t) &=& U_{}(t) \big[ \ket{\bar{x}_{a},\bar{p}_{a}}_\sigma\! \bra{\bar{x}_{a},\bar{p}_{a}} + \bar c e^{-i\bar\varphi}\ket{\bar{x}_{a},\bar{p}_{a}}_\sigma\! \bra{\bar{x}_{b},\bar{p}_{b}} \nn\\
		&+& \bar ce^{i\bar\varphi}\ket{\bar{x}_{b},\bar{p}_{b}}_\sigma\! \bra{\bar{x}_{a},\bar{p}_{a}} + \ket{\bar{x}_{b},\bar{p}_{b}}_\sigma\! \bra{\bar{x}_{b},\bar{p}_{b}}  \big] U^\dag(t). \nn\\ \label{rhobar}
	\end{eqnarray}
Here, we used $\bar x_a\equiv\bar x(x_g,x_a)$ etc.\ given by \Eqref{tracerrelation}, as well as
	\begin{eqnarray}
		\bar c &=&  \exp\!\left[ -\frac\alpha{(1+\alpha)^2} \left(\frac{x_D^2}{\sigma ^2} + \frac{\sigma^2p_D^2}{\hbar^2} \right)  \right], \label{cbar}
	\end{eqnarray}
which is smaller than one because of a reduction of coherences due to a \emph{measurement} performed by the colliding gas particle, and
	\begin{eqnarray}
		\bar\varphi &=& \frac{2\alpha(x_A p_D-x_D p_A) + (1-\alpha)p_gx_D - \alpha(1-\alpha)x_gp_D}{(1+\alpha)^2\hbar} \nn\\ \label{varphibar}
	\end{eqnarray}
is a phase shift induced by the collision. We further defined the average position $x_A=(x_a+x_b)/2$ and the position difference $x_D=(x_a-x_b)$ (analogous for momenta) of the two cohering wave packets.

Note that the phase shift $\varphi=\varphi(x_g,p_g)$ depends on the state of the colliding gas particle. If this state is random as in a thermal gas, the phase shift leads to phase averaging and therefore to a loss of coherences.

Rather than using a density operator representation in terms of Gaussian states, we found a Wigner function representation (see e.g.\ \cite{Busch}) more graphic to display coherences of cat states. Decoherence can then be discussed in terms of the vanishing of the oscillatory behavior of the Wigner function. In particular, we will compare the Wigner function without a collision, to the Wigner function with a collision. This way, we will obtain the \emph{`decoherence per collision'}, which can be multiplied with the collision rate \Eqref{rate}, to obtain the decoherence rate.

We wish to mention that a Wigner function description of quantum Brownian motion was already put forward previously in a very heuristic derivation~\cite{Halliwell}, as well as in a more precise approach, but limited to states of the tracer particle which are close to a thermal state~\cite{Hornberger3}. Both articles are concerned with the general form of a partial differential equation for the Wigner function, and do not study decoherence.

We will need the Wigner function corresponding to a density operator of the form
	\begin{eqnarray}
		\rho &=& \ket{x_a,p_a}_\sigma\!\bra{x_a,p_a} + ce^{-\imath\varphi} \ket{x_a,p_a}_\sigma\!\bra{x_b,p_b} \nn\\ 
		&+& ce^{\imath \varphi} \ket{x_b,p_b}_\sigma\!\bra{x_a,p_a} + \ket{x_b,p_b}_\sigma\!\bra{x_b,p_b}  . \label{state1}
	\end{eqnarray}
Here, $c$ is bound between zero and one and is a measure for the strength of the coherences between $\ket{x_a,p_a}_\sigma$ and $\ket{x_b,p_b}_\sigma$, and $\varphi$ determines the relative phase between these states.  In particular, an incoming tracer particle state $\ket{x_a,p_a}_\sigma+\ket{x_b,p_b}_\sigma$ before a collision corresponds to $c=1$ and $\varphi=0$.

Using standard methods~\cite{Busch}, the Wigner function corresponding to the density matrix \Eqref{state1} is easily found to be
	\begin{eqnarray}
		W_\rho(x',p') &=& \frac1{\pi\hbar}\exp\!\left[ -\frac{(x'-x_a)^2}{\sigma^2} \right] \exp\!\left[ -\frac{\sigma^2(p'-p_a)^2}{\hbar^2} \right] \nn\\
		&\hspace{-26mm}+& \hspace{-13mm}\frac1{\pi\hbar}\exp\!\left[ -\frac{(x'-x_b)^2}{\sigma^2} \right] \exp\!\left[ -\frac{\sigma^2(p'-p_b)^2}{\hbar^2} \right] \nn\\
		&\hspace{-26mm}+& \hspace{-13mm} \frac{2c}{\pi\hbar} \exp\!\left[ -\frac{(x'-x_A)^2}{\sigma^2} \right] \exp\!\left[ -\frac{\sigma^2(p'-p_A)^2}{\hbar^2} \right] \nn\\
		&\hspace{-26mm}\times& \hspace{-13mm} \cos\!\left[\varphi + \frac{x_Ap_D-p_Ax_D}{2\hbar} + x_D\frac{p_A-p'}{\hbar} - p_D\frac{x_A-x'}{\hbar}  \right]\!. \nn\\ \label{Wgeneral}
	\end{eqnarray}

\begin{figure}[t]
	\begin{center}
		\includegraphics[width=\linewidth]{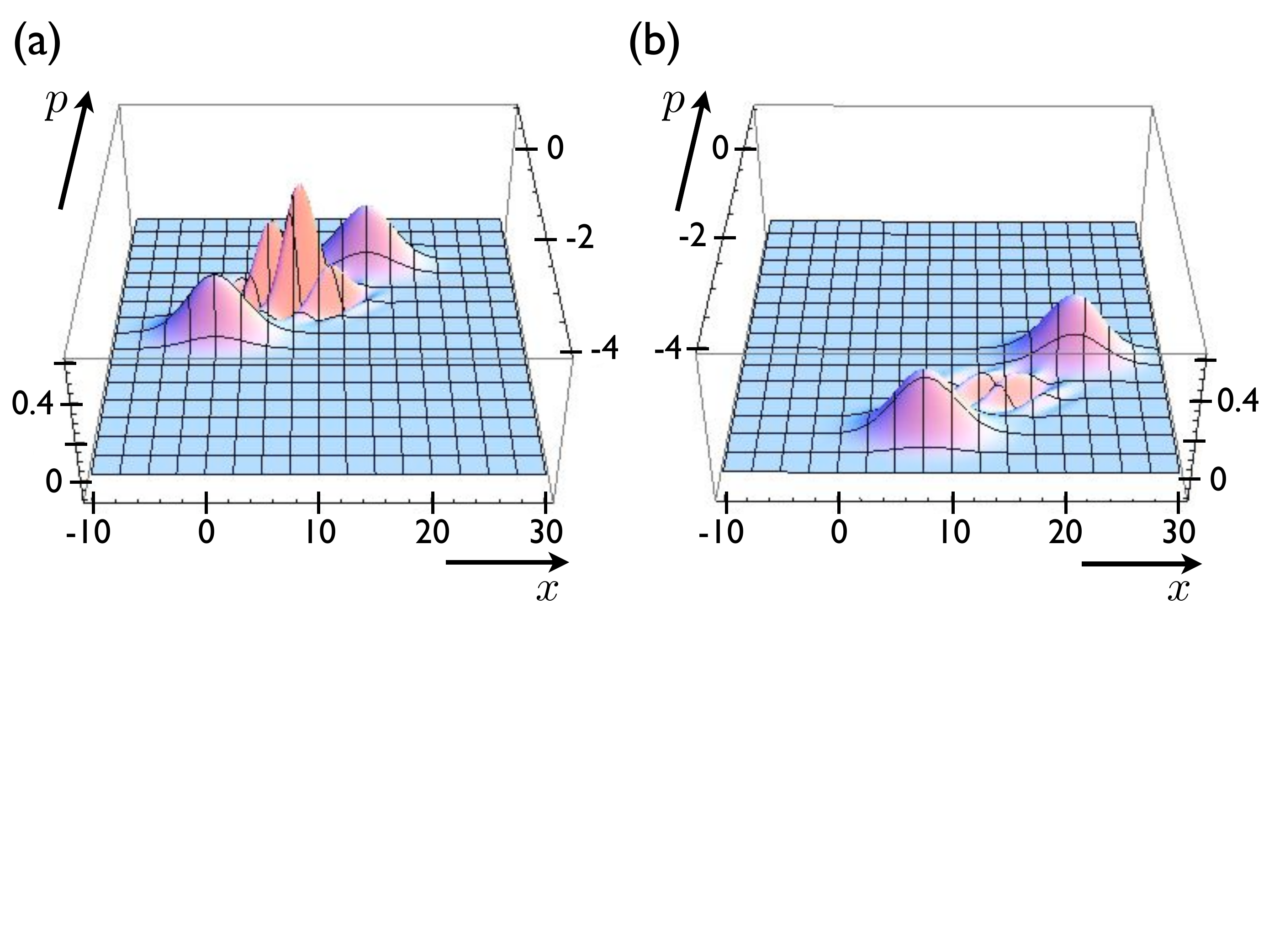}\vspace{-22mm}
		\caption{(color online) The Wigner function of the initial coherent superposition $\ket{\psi}=\ket{x_a,p_a}_\sigma+\ket{x_b,p_b}_\sigma$~(a), and of the state resulting from a collision with $\ket{x_g,p_g}_{\sigma_g}$~(b). Parameters are: $x_a=15,\;x_b=0,\;p_a=0,\;p_b=1.5,\;\sigma=4,\;m=1,\;\hbar=1,\;x_g=100,\;p_g=-1,$ and $\alpha=0.04$.}
	\label{Wcompare}
	\end{center}
	\end{figure}

As is well known, the strength of the coherences, indicated by oscillatory behavior of the Wigner function, does not change due to the unitary free evolution. Therefore, in the following, we will mostly use the interaction picture~\footnote{The time evolved Wigner function is obtained by replacing $x'$ by $(x'-p't/m)$ in \Eqref{Wgeneral}.}. We can then use the general formula \Eqref{Wgeneral}, with $c=1$ and $\varphi=0$, to plot the Wigner function of a state $\ket{\psi}=\ket{x_a,p_a}_\sigma+\ket{x_b,p_b}_\sigma$ without a collision in figure~\ref{Wcompare}~(a).

According to Eqs.~(\ref{rhobar}-\ref{varphibar}), a collision with a gas particle state $\ket{x_g,p_g}_{\sigma_g}$ results in
	\begin{eqnarray}
		(x_{a/b},p_{a/b},x_{A/D},p_{A/D}) &\to& (\bar x_{a/b},\bar p_{a/b},\bar x_{A/D},\bar p_{A/D}) \nn\\
		c=1 &\to& \bar c  \nn\\
		\varphi=0 &\to& \bar\varphi , \label{Wafter}
	\end{eqnarray}
where $\bar x_a=\bar x(x_g,x_a), etc$  are given by \Eqref{tracerrelation}, and $\bar c$ and $\bar\varphi$ by \Eqref{cbar} and \Eqref{varphibar}, respectively. The Wigner function after the collision is plotted in figure~\ref{Wcompare}~(b).

It is quite astonishing, that, despite choosing a gas particle with only four percent of the mass of the tracer particle, and, despite using a superposition of very close Gaussian wave functions, almost all coherences are lost after a single collision. If we had separated the initial Gaussians only slightly more, or had chosen only a slightly heavier gas particle, the coherences would be not visible at all, because $\bar c$ in \Eqref{cbar} decreases exponentially with these parameters. This observation is independent of the initial momentum and position of the colliding gas particle, as well as whether the Gaussian wave functions are separated predominantly in position or momentum. It is therefore fair to say, that, unless the gas particle is much lighter than the tracer particle, the decoherence rate equals the collision rate.

This result becomes even more pronounced, if we average over different initial gas particle positions and momenta, and we will study this effect in the following section for extremely light gas particles, for which the decoherence per collision due to information exchange of the colliding particles, $(1-\bar c)$, will be small.

\section{Collisional decoherence for light gas particles\label{secII}}

In this section, we discuss the more interesting situation, when a collision only partially destroys the coherences, which is the case if the colliding gas particle is extremely light (mass ratios much smaller than one percent, as we will see). Then, we need a more quantitative measure of the decoherence process, which we will develop below.

\begin{figure}[t]
\begin{center}
	\includegraphics[width=\linewidth]{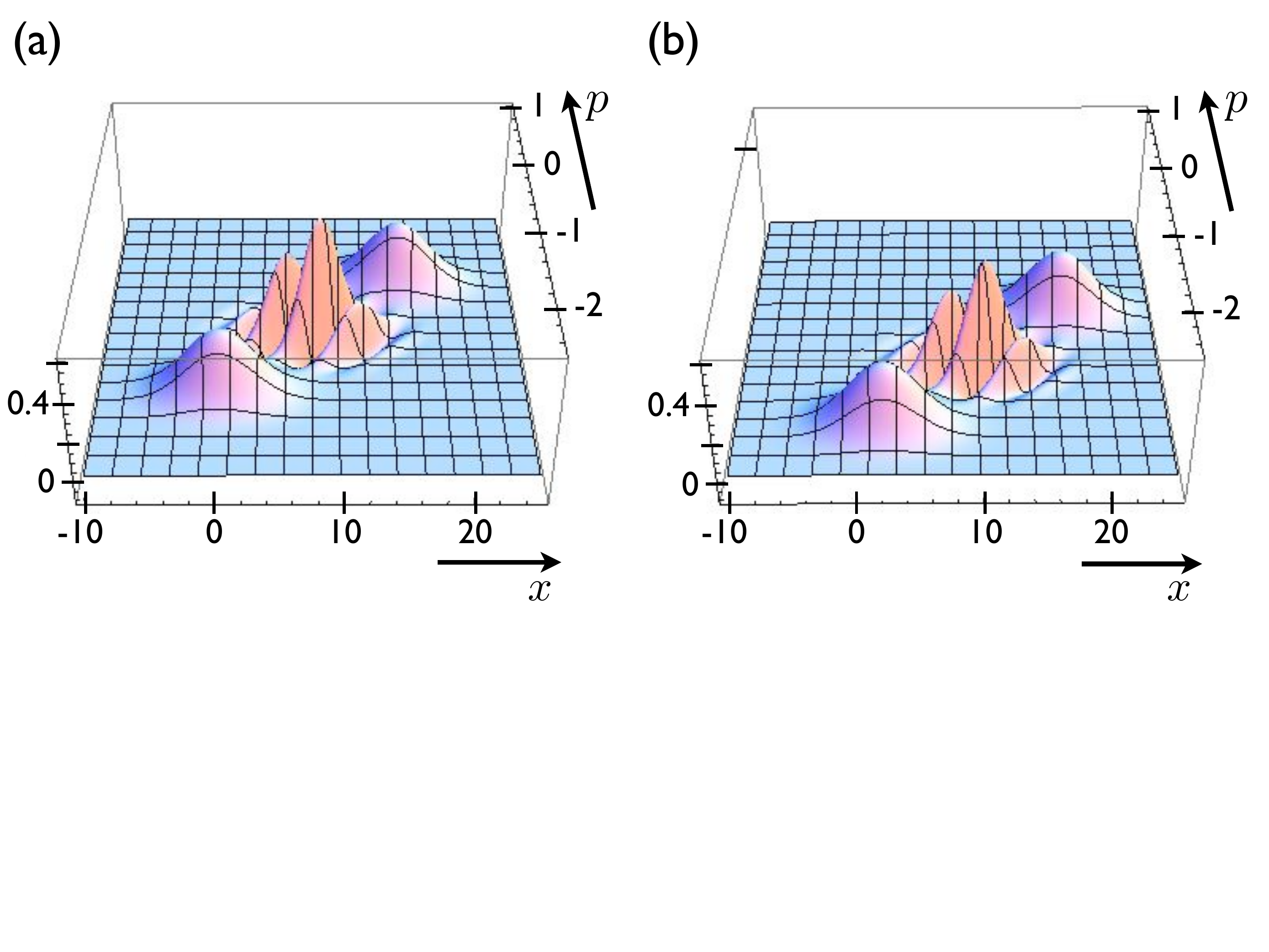}\vspace{-22mm}
\caption{(color online) As in figure~\ref{Wcompare}, but with different parameters for the gas particle: $p=-0.2,\;x_g=500,$ and $\alpha=0.002$.}
\label{Wcompare2}
\end{center}
\end{figure}

Figure~\ref{Wcompare2} shows how the Wigner function of a superposition state changes due to a collision with a light gas particle. The coherences after the collision are still well resolved. The reason is that the measurement performed by the colliding gas particle is so imprecise, that it can not distinguish between the two Gaussian wave functions of the tracer particle~\footnote{This can also be seen from the Kraus operators which represent the transformation of the tracer particle state due to a collision. The Kraus operators are derived in~\cite{Kamleitner} and consist of the Glauber displacement operator with a small correction representing a weak phase space measurement.}.

The main effect of a collision with a light gas particle is a shift of the entire Wigner function in phase space. Of particular interest is that the oscillating part just shifts from $(x_A,p_A)$ to $(\bar x_A,\bar p_A)$, without acquiring an additional phase shift (i.e., the relative heights of the oscillating peaks do not change). This feature can be explained by looking at the argument of the cosine in \Eqref{Wgeneral}. The sum of the first two terms does not change in a collision, because $(x_Ap_D-p_Ax_D)/(2\hbar)$ equals $(\bar x_A\bar p_D-\bar p_A\bar x_D)/(2\hbar)+\bar\varphi$, with $\bar\varphi$ taken from \Eqref{varphibar}. The third and forth term account for a phase shift corresponding to the shift of the Gaussian, i.e.\ $x_A-x'\to\bar x_A-x'$ and $p_A-p'\to\bar p_A-p'$.


Of course, if the gas particle state $\ket{x_g,p_g}$ is taken from a thermal gas, we have to average over all gas particle momenta $p_g$, weighted by the Maxwell-Boltzmann distribution $\mu_T(p_g)$, as well as over all the gas particle positions $x_g$ which can reach the tracer particle in a given time interval $(0,t)$. Because each possible combination of $x_g$ and $p_g$ results in a different shift of the Wigner function in phase space, it is clear that this procedure strongly suppresses the oscillations. This is the decoherence effect we referred to as `phase averaging' in the introduction. It can suppress coherences quickly, even if the measurements which the gas particles perform are very weak. 

Let us close this preliminary discussion with a note regarding the collision rate. In the limit of a small mass ratio $\alpha$, the tracer particle will be very localized compared to the gas particles, in both, position and velocity, as is evident from \Eqref{sigmarel}. Therefore, we do not have to use the full rate operator formalism developed in~\cite{Kamleitner}, but we can simply assume that the tracer particle is reasonably localized somewhere near the origin, and a gas particle collides with the tracer particle during a time interval $(0,t)$ exactly if 
	\begin{equation}
		0<-x_gm_g/p_g<t \label{collbed}
	\end{equation}
is satisfied.

\subsection{Position decoherence}

\begin{figure}[t]
	\includegraphics[width=\linewidth]{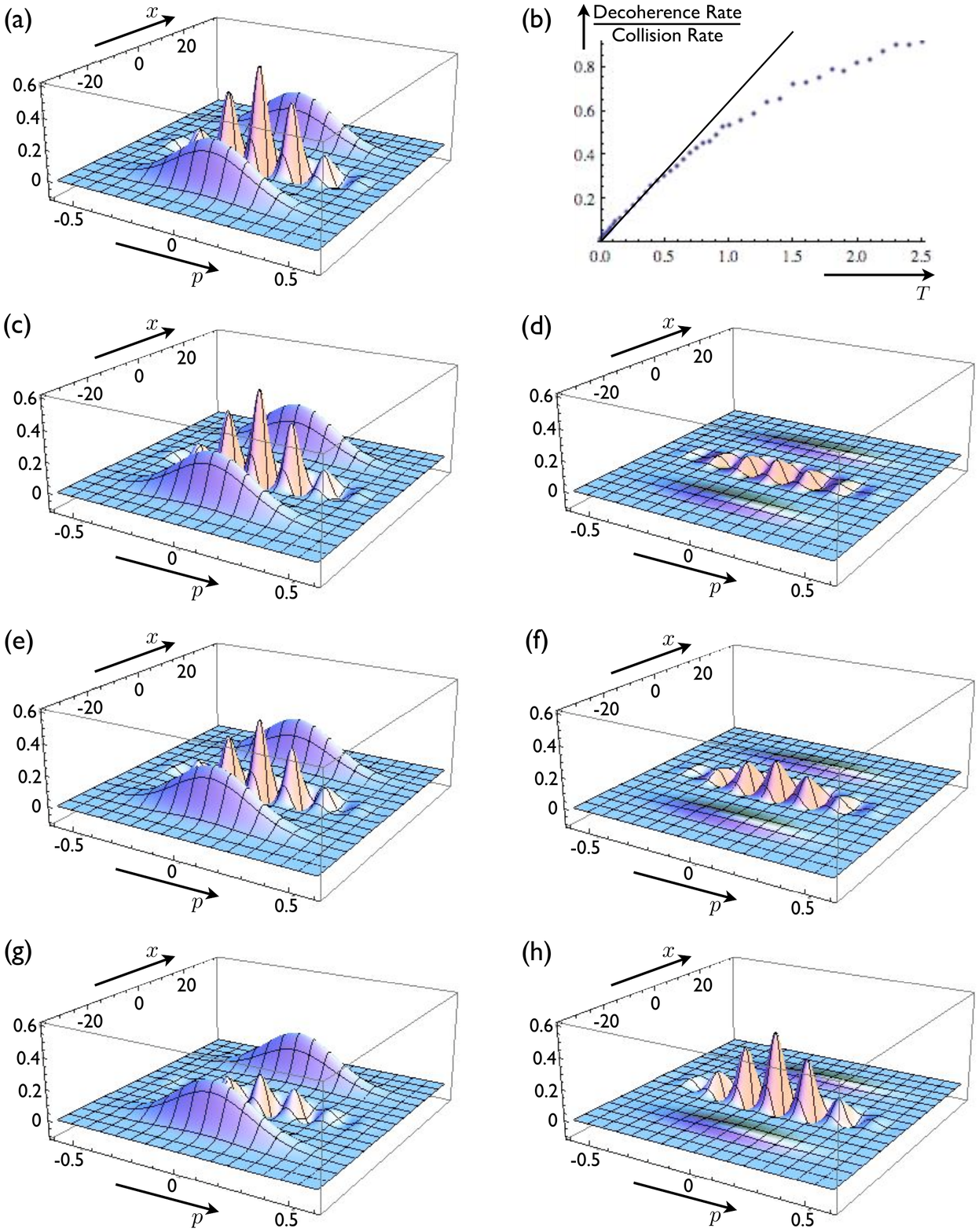}\vspace{-11mm}
	\caption{(color online) (a): The Wigner function for the tracer particle before a collision. (c), (e), (g): The average Wigner function after one collision with a gas particle at temperature $T=0.2,\;0.5$, and $1.5$, respectively. (d), (f), (h): The change of the Wigner function due to a collision at temperature $T=0.2,\;0.5$, and $1.5$, respectively. (b): The relative change of the Wigner function at the origin due to a collision. This serves as a quantitative measure of the `decoherence per collision'. The solid line is the first order expansion in temperature, \Eqref{qscz}. Parameters are: $x_a=20,\;x_b=-20,\;p_a=0,\;p_b=0,\;\sigma=4,\;m=1,\;\hbar=1,\;k_B=1,\;t=20,$ and $\alpha=0.0001$.}
	\label{Wpos}
\end{figure}

To study position decoherence, we consider an initial tracer particle state $\ket{x_a,p}_\sigma+\ket{x_b,p}_\sigma$. An example of the corresponding Wigner function is plotted in figure~\ref{Wpos}~(a). The Wigner function after a collision with a gas particle state $\ket{x_g,p_g}_{\sigma_g}$ is obtained following section~\ref{medi}, but to study the decoherence due to a thermal gas, we average over different initial positions $x_g$ and momenta $p_g$ of the gas particle.

%

The Wigner function $W_{\rho_1}$ after a collision, averaged over 200 pairs $(x_g,p_g)$ taken from an appropriate distribution (see below), is shown in figure~\ref{Wpos}~(c), (e), and (g), for the respective temperatures $T=0.2,\;0.5$, and $1.5$. The corresponding changes of the Wigner function from the initial one (figure~\ref{Wpos}~(a)) are shown figure~\ref{Wpos}~(d), (f), and (g). 

As a quantitative measure of coherence, we use the hight of the maximum peak of the oscillations. In particular, the relative change of the maximum peak serves as `decoherence per collision', and is plotted over the temperature $T$ in figure~\ref{Wpos}~(b). 

In the following, we derive an expression for the `decoherence per collision'. In the limit of a light gas particle we can use $\bar c\approx1$, and concentrate on the cosine of the Wigner function \Eqref{Wgeneral} 
	\begin{eqnarray}
		\cos\!\left[\varphi - \frac{p\,x_D}{2\hbar} + x_D\frac{p-p'}{\hbar} \right],  \label{Wosci}
	\end{eqnarray}
where we used $p_D=0$ and $p_A=p$ for our choice of the initial cat state.

As noted earlier, a collision does not change the sum $\varphi-px_D/(2\hbar)$. Further, we use $1+\alpha\approx1$ in the limit of light gas particles, which leads to $\bar x_D\approx x_D$ and $\bar p\approx p+2p_g$. Therefore, we find the oscillating term of the Wigner function $W_{\rho_1}$ after a collision by averaging over
	\begin{eqnarray}
		\cos\!\left[\varphi - \frac{p\,x_D}{2\hbar} + x_D\frac{p+2p_g-p'}{\hbar} \right]  \label{Wosci}
	\end{eqnarray}
In particular, we use the reduction of the maximum of the oscillations as a quantitative measure of decoherence. Because this maximum before a collision is at $p'$ given by $\varphi -p\,x_D/(2\hbar) + x_D(p-p')/\hbar=0$, we substitute this into \Eqref{Wosci}, and the coherences after one collision are then obtained by averaging over
	\begin{eqnarray}
		\cos\!\left(\frac{2x_D\,p_g}{\hbar}\right).  \label{Wosci2}
	\end{eqnarray}
We see that the relative phase $2p_gx_D/\hbar$ added by the collision depends solely on the momentum of the colliding gas particle, and we will need the momentum probability distribution of the colliding gas particles to perform the averaging over \Eqref{Wosci2}.

According to \Eqref{convex}, the probability density of finding a gas particle in the state $\ket{x_g,p_g}_{\sigma_g}$ is given by $n_g\mu_T(p_g)$, where $n_g$ is the particle density of the gas. Knowing that a gas particle of momentum $p_g$ collides with the tracer particle exactly if the position $x_g$ satisfies \Eqref{collbed}, we can write down the normalized momentum probability distribution of a colliding gas particle
	\begin{eqnarray}
		C(p_g)&=& \frac{|p_g|}{2m_gk_BT}\exp\!\left(-\frac{p_g^2}{2m_gk_BT}\right). \label{gjw}
	\end{eqnarray}
Note that this distribution does not depend on the length of the considered time interval $(0,t)$. This will lead to a `decoherence per collision' (and therefore to a decoherence rate) which is independent of the considered time interval, as should be expected in a Markovian process.

We finally find for the coherences after one collision with a thermal gas particle
\begin{widetext}
	\begin{eqnarray}
		\left\langle  \cos\!\left(\frac{2x_D\,p_g}{\hbar}\right)\! \right\rangle_{\!C(p_g)} &=& \int_0^\infty \dd p_g\frac{p_g}{m_gk_BT} \exp\!\left( \frac{-p_g^2}{2m_gk_BT}\right) \cos\!\left(\frac{2x_D\,p_g}{\hbar}\right)\label{surprise1} \\
		&=& 1-\frac{2x_D}{\hbar} \int_0^\infty \dd p_g\, \exp\!\left( \frac{-p_g^2}{2m_gk_BT}\right) \sin\!\left(\frac{2x_D\,p_g}{\hbar}\right)\!,  \label{surprise}
	\end{eqnarray}
where we used integration by parts. We deduce for the `decoherence per collision' of position superposition states
	\begin{eqnarray}
		\frac{\mbox{Decoherence}}{\mbox{Collision}}&=& \frac{2x_D\sqrt{2m_gk_BT}}{\hbar} \int_0^\infty \dd u\, e^{-u^2} \sin\!\left(\frac{2x_D\sqrt{2m_gk_BT}}{\hbar}u\right)\!, \label{surprise2}
	\end{eqnarray}
or for the decoherence rate
	\begin{eqnarray}
		D_x &=& \frac{4x_Dn_gk_BT}{\sqrt\pi \hbar} \int_0^\infty \dd u\, e^{-u^2} \sin\!\left(\frac{2x_D\sqrt{2m_gk_BT}}{\hbar}u\right)\!.  \label{surprise3}
	\end{eqnarray}
\end{widetext}

An interesting limit to consider is $2x_D\sqrt{2m_gk_BT}\ll\hbar$, in which the `decoherence per collision' is small. This situation corresponds either to a small position separation $x_D$, or to a small average momentum transfer $\sqrt{2m_gk_BT}$, and was studied in the seminal work of Joos and Zeh~\cite{Joos}. The integrals in \Eqref{surprise2}-(\ref{surprise3}) can then be carried out and we find
	\begin{eqnarray}
		\frac{\mbox{Decoherence}}{\mbox{Collision}}&=&\frac{4m_gk_BT}{\hbar^2}x_D^2. \label{qscz}\\
		D_x&=& \frac{8n _g\sqrt{m_g}(k_BT)^{3/2}}{\sqrt{2\pi}\hbar^2}x_D^2.
	\end{eqnarray}
The decoherence rate $D_x$ agrees up to some constant with the one found by \cite{Joos} for three dimensional collisions.

\begin{figure}[t]\vspace{0cm}
	\includegraphics[width=\linewidth]{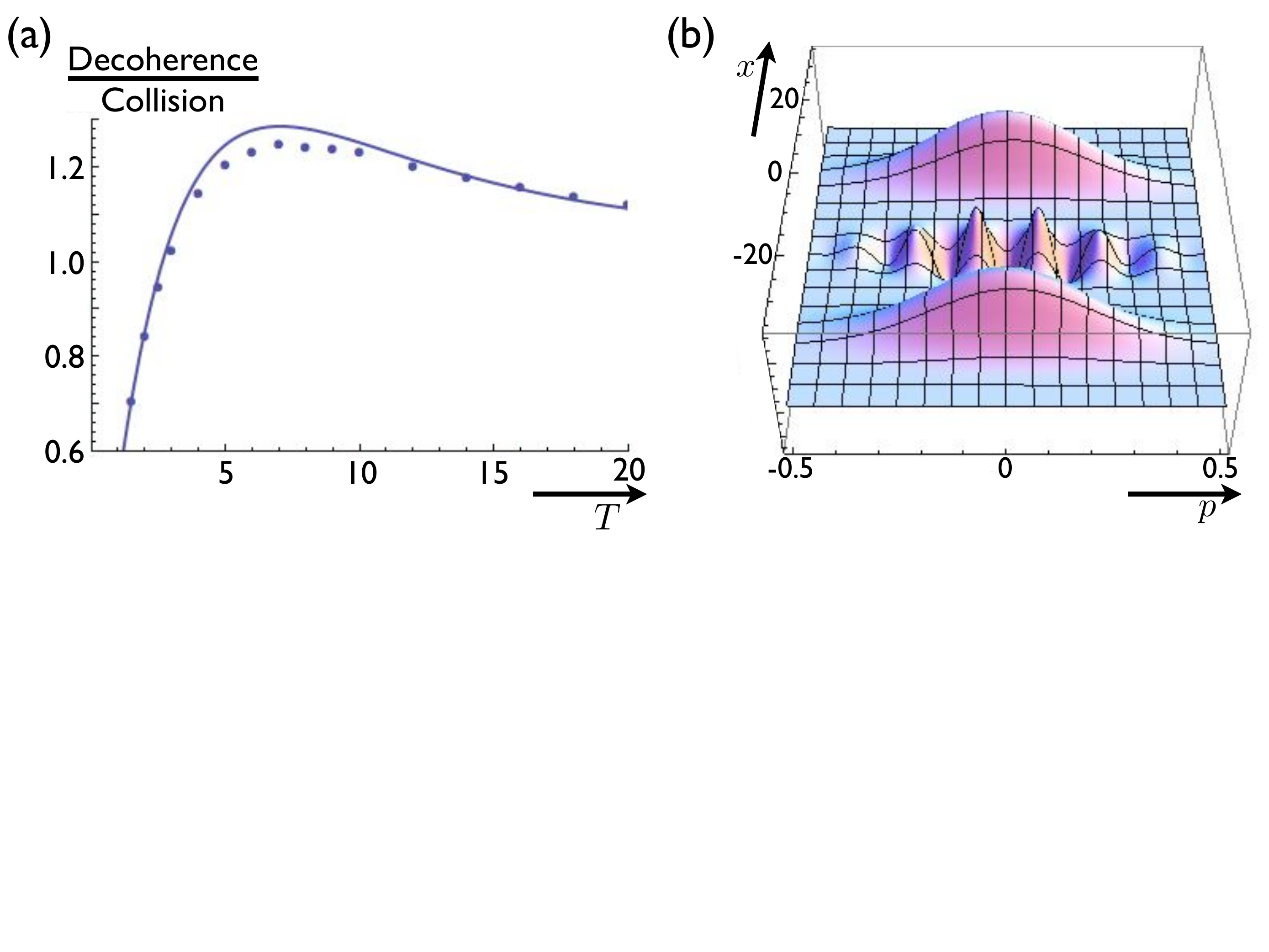} \vspace{-32mm}
	\caption{(color online) (a): The `decoherence per collision' plotted over the temperature, for parameters as in figure~\ref{Wpos}, but for larger temperatures. The dots are from the numerical Wigner function at the origin, and the line is from \Eqref{surprise}. The small discrepancy is due to the neglected spreading of the Gaussians due to the averaging process. \newline (b): The Wigner function after a collision at temperature $T=7$ shows indeed oscillations in opposite phase to the initial Wigner function (see figure~\ref{Wpos}~(a)). This leads to a decoherence rate which exceeds the collision rate.}
	\label{Wsurprise}
\end{figure}

The `decoherence per collision' \Eqref{surprise2} is plotted over temperature in figure~\ref{Wsurprise}~(a) for the same parameters as in figure~\ref{Wpos}~(b), but for higher temperatures. It might come as a surprise, that the decoherence rate exceeds the collision rate for $x_D\sqrt{2m_gk_BT}/\hbar \gtrsim 1$. The reason is that in this regime, the Wigner function after a collision shows oscillations, which are out of phase with the oscillations of the initial Wigner function, as shown in figure~\ref{Wsurprise}~(b). Therefore, if we write down the actual (interaction picture) Wigner function after some small time $t$ as 
	\begin{eqnarray}
		W_{\rho(t)} &=& (1-Rt)W_{\rho_0}+RtW_{\rho_1},
	\end{eqnarray}
where $R$ is the collision rate \Eqref{rate}, and $\rho_0$ and $\rho_1$ are the density operators corresponding to no collision and to one collision, respectively, then the oscillations in $W_{\rho_0}$ and $W_{\rho_1}$ interfere destructively. These out-of-phase coherences in turn can be understood by noting that the momentum distribution of the colliding particles, \Eqref{gjw}, is not peaked at $p_g=0$, but rather at $p_g=\pm\sqrt{2m_gk_BT}$.

Having found the `decoherence per collision' due to phase averaging, we can draw a quantitative comparison with the `decoherence per collision' due to information exchange, which is $1-\bar c\approx \alpha x_D^2/\sigma^2$. Because of the first inequality of (\ref{limitation}), this decoherence effect is indeed negligible if the density and temperature of the gas are such, that it can be considered an ideal Boltzmann gas. This is also true in the regime discussed in section~\ref{medi}, because phase averaging is sufficient to remove any coherences in a single collision.

We note that it is often stated in the literature \cite{Joos,review} that, if the separation $x_D$ of two interfering wave packets is larger than the thermal wave length $\Lambda=\hbar/\sqrt{2\pi m_g k_BT}$ of the gas, then a colliding gas particle can distinguish between the two interfering wave packets, therefore removing their coherences. In finding that the decoherence rate is about the collision rate if $x_D \gtrsim \sqrt{\pi}\Lambda$, we confirm the latter part of this statement, but we also show that the loss of coherence is by no means related to a measurement performed by the gas particle, but due to \emph{classical} phase averaging resulting from the randomness of the momentum transfer.

\subsection{Momentum decoherence}

In this subsection, we will show that the decoherence of momentum superposition states is not a direct process. Instead, two coherent wave packets with momentum separation $p_D$ will, after some time, acquire a position separation $x_D=p_Dt/m$, which then leads to position decoherence. Any direct momentum decoherence will turn out to be negligible in a high temperature and low density gas.

\begin{figure}[t]
	\includegraphics[width=\linewidth]{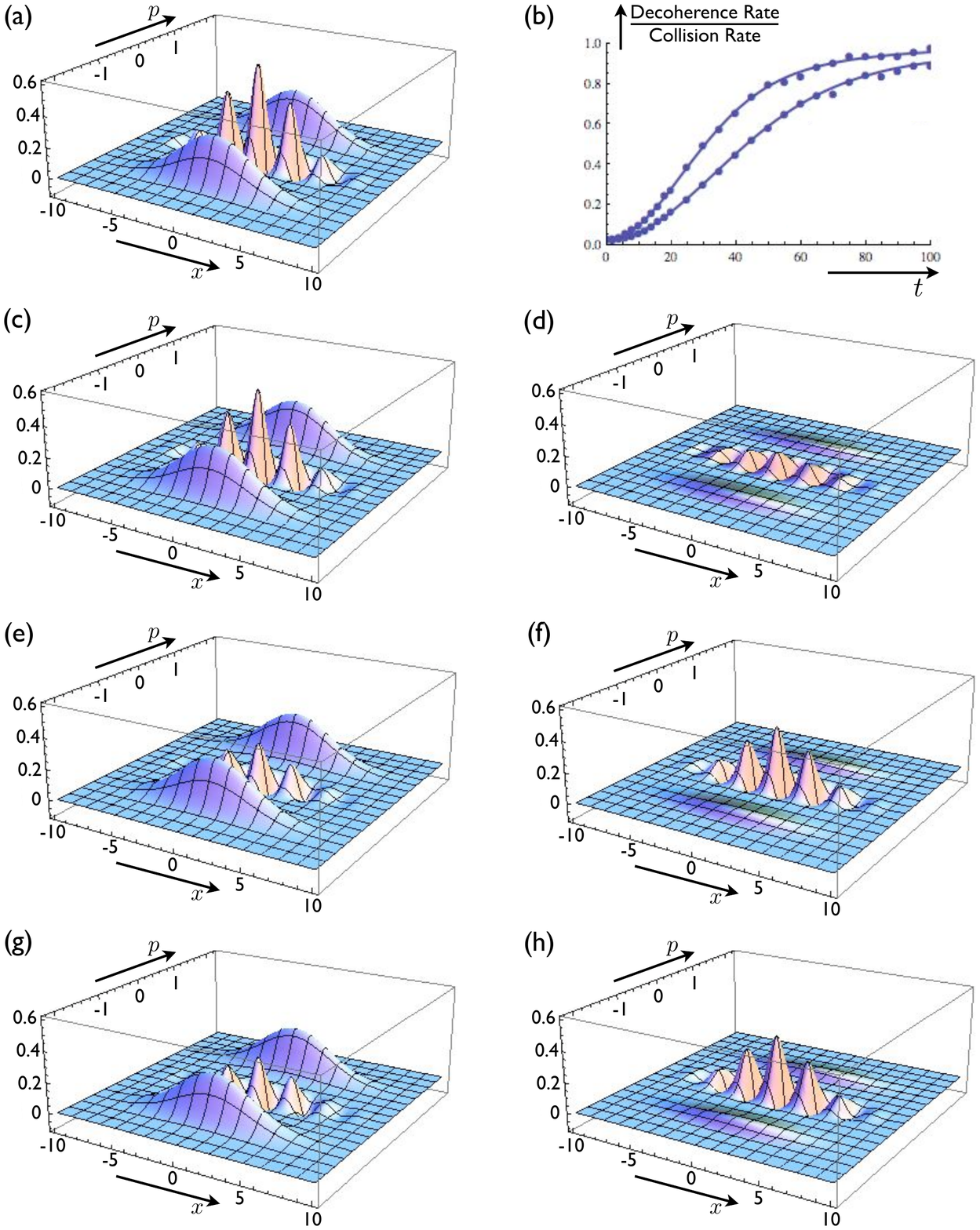}\vspace{-12mm}
	\caption{(color online) (a): The Wigner function for the tracer particle before a collision. (c), (e), (g): The average Wigner function after one collision with a gas particle within a time interval $(0,t)$ at temperature $T$, with $t=20,\;T=0.5$ in (c);  $t=50,\;T=0.5$ in (e);  $t=20,\;T=3$ in (g). (d), (f), (h): The corresponding change of the Wigner function due to a collision. (b): The relative change of the Wigner function at the origin due to a collision as function of the time interval for $T=0.5$ as well as $T=1$. Parameters are: $x_a=0,\;x_b=0,\;p_a=1.2,\;p_b=-1.2,\;\sigma=4,\;m=1,\;\hbar=1,\;k_B=1,$ and $\alpha=0.0001$.}
	\label{Wmom}
\end{figure}

We consider the initial tracer particle state $\ket{x,p_a}_\sigma+\ket{x,p_b}_\sigma$, whose Wigner function is plotted in figure~\ref{Wmom}~(a). Again, the main source of decoherence will be phase averaging. Contrary to the previous subsection, where the relative phase of the two Gaussian wave packets after a collision depended on the initial momentum of the colliding gas particle, for momentum decoherence this phase depends on the initial position of the colliding gas particle~\footnote{This can be spelled out in a more intuitive way by saying that the acquired relative phase depends on the actual time of collision $t'\in(0,t)$, multiplied by the momentum of the colliding gas particle.}. As the variation of the initial position of a colliding gas particle is increased, the more `decoherence per collision' we will find. For a given gas particle momenta $p_g$, the gas particle position can be anywhere within the interval $(-p_gt/m_g,0)$. Therefore, the `decoherence per collision' will not only depend on the temperature $T$ (for the distribution of $p_g$), but also on the considered time interval. In figure~\ref{Wmom}~(c) - (h), the effects of a collision on the Wigner function is shown for different temperatures and time intervals. The dependence of the `decoherence per collision' for low temperatures and short times turns out to be linear in temperature and quadratic in time, as shown in~figure~\ref{Wmom}~(b). As a result, it is not possible to define a time independent decoherence rate for momentum superpositions. We will provide a physical interpretation at the end of this subsection, and first give a mathematical explanation of these results.

For this purpose, we consider again the oscillating cosine within the Wigner function \Eqref{Wgeneral}
	\begin{eqnarray}
		\cos\!\left[\varphi + \frac{x\,p_D}{2\hbar}  - p_D\frac{x-x'}{\hbar}  \right]\!, \label{tfcd}
	\end{eqnarray}
at its maximum $\varphi +xp_D/(2\hbar)-p_D(x-x')/\hbar=0$. As discussed before, a collision does not change the sum $\varphi +xp_D/(2\hbar)$ in the cosine, and we only have to consider the change of $x$ to $x+2\alpha x_g$ in the last term of the cosine. Therefore, we find the `decoherence per collision' by averaging over $\cos(2\alpha x_g\,p_D/\hbar)$. Here, we need the normalized probability distribution $\widetilde C(x_g)$ of the initial position of the colliding gas particle, which is obtained from the probability density $n_g\mu_T(p_g)$ by integration over all $p_g$, for which a gas particle with position $x_g$ can reach the tracer particle (i.e.\ $-x_g\gtrless p_gt/m_g \gtrless \mp \infty$, where the upper sign is for positive $x_g$)
	\begin{eqnarray}
		\widetilde C(x_g) &\propto& \frac{n_g}{\sqrt{2\pi m_gk_BT}} \int_{|x_g|m_g/t}^\infty \dd p_g \exp\!\left(-\frac{p_g^2}{2m_gk_BT}\right)\!.\nn
	\end{eqnarray}
The distribution is normalized either by integration over $x_g$, or directly by dividing by the collision probability $Rt$. After substituting $u=p_g/\sqrt{2m_gk_BT}$ we find
	\begin{eqnarray}
		\widetilde C(x_g) &=& \frac{\sqrt{m_g}}{t\sqrt{2k_BT}} \int_{\frac{|x_g|\sqrt{m_g}}{t\sqrt{2k_BT}}}^\infty \dd u\,e^{-u^2},
	\end{eqnarray}
and therefore
\begin{widetext}
	\begin{eqnarray}
		\left\langle  \cos\!\left(\frac{2\alpha p_D\,x_g}{\hbar}\right) \right\rangle_{\!\widetilde C(x_g)} &=& \frac{\sqrt{m_g}}{t\sqrt{2k_BT}} \int_{-\infty}^\infty \dd x_g\,\cos\!\left(\frac{2\alpha p_D\,x_g}{\hbar}\right) \int_{\frac{|x_g|\sqrt{m_g}}{t\sqrt{2k_BT}}}^\infty \dd u\,e^{-u^2} \nn\\
		&=& \frac{m\hbar}{t\sqrt{2m_gk_BT}p_D}\int_0^\infty \dd u\, e^{-u^2} \sin\!\left( \frac{2t\sqrt{2m_gk_BT}p_D}{m\hbar}u \right) \!. \label{pjrd}
	\end{eqnarray}
Because this function represents the coherences after one collision, we have to subtract it from unity to obtain the `decoherence per collision'
	\begin{eqnarray}
		\frac{\mbox{Decoherence}}{\mbox{Collision}} &=& 1- \frac{m\hbar}{t\sqrt{2m_gk_BT}p_D}\int_0^\infty \dd u\, e^{-u^2} \sin\!\left( \frac{2t\sqrt{2m_gk_BT}p_D}{m\hbar}u \right) \qquad\quad \label{kfur} \\
		&\approx& \frac{4m_gk_BT}{3\hbar^2}\left(\frac{tp_D}{m}\right)^2 .\label{kfur2}
	\end{eqnarray}
\end{widetext}
The solid lines in figure~\ref{Wmom}~(b) are taken from \Eqref{kfur}, and agree well with the data (dots) obtained from the numerical Wigner functions directly. The approximation in \Eqref{kfur2} is valid if the `decoherence per collision' is small. 

Similar to position decoherence, we see from \Eqref{kfur2} that momentum decoherence due to information exchange ($1-\bar c\approx \alpha \sigma^2p_D^2/\hbar^2$) is negligible (unless for temperatures and times so small, that relation~(\ref{colltime}) is violated). Hence, we established that also momentum decoherence is due to phase averaging.

At first, the increase of the decoherence rate with the considered time interval seems to be at odds with the uniformity in time in the following sense: If we split a time interval $(0,t)$ into sub intervals $(0,t/N),\;(t/N,2t/N),\dots,(t-t/N,t)$, the decoherence rate of the entire interval should be the averaged decoherence rate of all the sub intervals. If we now assume that the decoherence rate for each sub interval is the same (``uniformity in time''), we would be lead to the conclusion that the decoherence rate for the interval $(0,t)$ equals the decoherence rate for the subinterval $(0,t/N)$, clearly contradicting \Eqref{kfur} and \Eqref{kfur2}.

In the above argument, we made the  following conceptual error: by assuming the same decoherence rate for each sub interval, we implied the same initial cat state at the beginning of each sub interval. But instead, by the time the $n$-th sub interval starts, the cat state evolved to $U(nt/N)(\ket{x,p_a}_\sigma+\ket{x,p_b}_\sigma)$, and acquired a position separation $x_D=p_D(nt/N)/m$. Therefore, we have to add position decoherence for all but the first sub intervals. 

In the limit $N\to\infty$, there is no momentum decoherence in the infinitely small sub intervals at all, but instead, a continuously increasing position decoherence according to \Eqref{surprise2}, with $x_D(t')=p_Dt'/m$. Indeed, substituting this time dependent position separation into \Eqref{surprise2} and averaging over all times $t'\in(0,t)$, one exactly recovers \Eqref{kfur}~\footnote{This is easier shown by recovering \Eqref{pjrd} from \Eqref{surprise1}.}.

In other words, the decoherence which an initial cat state $\ket{x,p_a}+\ket{x,p_b}$ experiences during a time interval $(0,t)$ is perfectly explained by position decoherence of the evolving state $U(t')(\ket{x,p_a}+\ket{x,p_b})$. This leads us to the physical interpretation that momentum decoherence is not a direct process, but results indirectly from position decoherence due to position separation which the tracer particle acquires over time.

\section{Conclusions from the study of decoherence}

We showed previously~\cite{Kamleitner} that in one dimensional collisional decoherence, the quantity measured by a colliding gas particle does not only depend on physical parameters like density and temperature, but also on the choice of decomposition \Eqref{convex} of the density operator of a thermal gas particle. It is therefore reassuring to find in this article, that measurement effects as a source of decoherence are negligible in the high temperature and low density limit, where \Eqref{convex} is valid. The reason is that measurement effects are small compared to phase averaging effects, which arise from a random relative phase added to a superposition state during the collision process.

We further arrive at a neat interpretation of the decoherence process of a superposition of two Gaussians wave packets. The decoherence due to a collision depends on the position separation of the two Gaussian wave packets at the time of the collision. In contrast, there is no direct decoherence due to the momentum separation $p_D$ of the two coherent wave packets. Instead, over time, the momentum separation changes the position separation according to $x_D(t)=x_D+tp_D/m$. This leads to an indirect influence of $p_D$ on the decoherence rate, which, if there is no initial position separation, is described by \Eqref{kfur}.

Further work is required to see whether this drastic change in the understanding of the collisional decoherence process also applies to three dimensional systems.

\acknowledgments{I would like to thank Dr.\ James Cresser for many helpful discussions.

Most of this work was funded under the MQRES scheme of Macquarie University.}


\begin{thebibliography}{99}

\bibitem{Ghirardi} G.C. Ghirardi, A. Rimini, and T. Weber, Phys. Rev. D \textbf{34}, 470 (1986).
\bibitem{Zurek} W.H. Zurek, Rev. Mod. Phys. \textbf{75}, 715 (2003).
\bibitem{Joos} E. Joos and H.D. Zeh, Z. Phys. B: Condens. Matt. \textbf{59}, 223 (1985).
\bibitem{Gallis} M.R. Gallis and G.N. Fleming, Phys. Rev. A \textbf{42}, 38 (1990).
\bibitem{Hornberger} K. Hornberger and J. E. Sipe, Phys. Rev. A \textbf{68}, 012105 (2003).
\bibitem{Dodd} P.J. Dodd and J.J. Halliwell, Phys. Rev. D \textbf{67}, 105018 (2003).

\bibitem{Diosi} L. Di\'osi, Europhys. Lett. \textbf{30}, 63 (1995).
\bibitem{Vacchini} B. Vacchini, Phys. Rev. Lett. \textbf{84}, 1374 (2000); B. Vacchini, Phys. Rev. E \textbf{63}, 066115 (2001).
\bibitem{main} S. M. Barnett and J. D. Cresser, Phys. Rev. A \textbf{72}, 022107 (2005).
\bibitem{Hornberger2}  K. Hornberger, Phys. Rev. Lett. \textbf{97}, 060601 (2006).
\bibitem{Breuer2} H.-P. Breuer and B. Vacchini, Phys. Rev. E \textbf{76}. 036706 (2007).

\bibitem{Hornberger4} M. Busse, P. Pietrulewicz, H.-P. Breuer, K. Hornberger, quant-ph, arXiv:10050365 (2010).
\bibitem{Diosi2} L. Di\'osi, Phys. Rev. A \textbf{80} 064104 (2009).
\bibitem{Kamleitner} I. Kamleitner and J. Cresser, Phys. Rev. A \textbf{81}, 012107 (2010).
\bibitem{review} B. Vacchini and K. Hornberger, Phys. Rep. \textbf{478}, 71 (2009).

\bibitem{Schmuser} F. Schm\"user and D. Janzing, Phys Rev A \textbf{73}, 052313 (2006).
\bibitem{lecture} K. Hornberger, Lect. Notes Phys. \textbf{768}, 221 (2009).
\bibitem{Halliwell} J.J. Halliwell, J. Phys. A: Math. Theor. \textbf{40}, 3067 (2007).
\bibitem{Hornberger3} B. Vacchini and K. Hornberger, Eur. Phys. J. Special Topics \textbf{151}, 59 (2007).
\bibitem{Busch} P. Busch, M. Grabowski, and P.J. Lahti, \emph{Operational Quantum Physics}, Springer-Verlag Berlin Heidelberg, 1995








%
%
%

\end{thebibliography}
\end{document}